\documentclass{aa}
\usepackage{times,natbib,graphicx}
\usepackage[english]{babel}
\begin{document}

\title{An ISO/SWS study of the dust composition around S
  stars\thanks{based on observations obtained with ISO, an ESA project
    with instruments funded by ESA Member states (especially the PI
    countries: France, Germany, the Netherlands and the United
    Kingdom) and with the participation of ISAS and NASA.}}

\subtitle{A novel view of S-star dust}

\author{
  S. Hony\inst{1},
  A. M. Heras\inst{2},
  F. J. Molster\inst{3}
  K. Smolders\inst{4}
  }
\authorrunning{Hony et al.}

\offprints{S. Hony (sacha.hony@cea.fr)} 
\institute{
  Laboratoire AIM, CEA/DSM - CNRS - Universit\'e Paris Diderot,
  DAPNIA/Service d'Astrophysique, B\^at. 709, CEA-Saclay, F-91191
  Gif-sur- Yvette C\'edex, France
  \and
  Research and Scientific Support Department-ESA/ESTEC, P.O. Box 299,
  2200 AG Noordwijk, The Netherlands
  \and
  NOVA, P.O. Box 9513, 2300 RA Leiden,The Netherlands
  \and
  Instituut voor Sterrenkunde, K.U. Leuven, Celestijnenlaan 200D,
  B-3001 Leuven, Belgium }

\date{received \today; accepted date}

  \abstract
  {}
  {We investigate the composition of the solid-state materials in the
    winds around S-type AGB stars. The S stars produce dust in their
    wind that bears a resemblance to the dust produced in some O-rich
    AGB stars. However, the reported resemblance is mostly based on
    IRAS/LRS spectra with limited spectral resolution, sensitivity,
    and wavelength coverage.}
  {We investigate the dust composition around S stars using ISO/SWS
    data that surpass the previous studies in terms of spectral
    resolution and wavelength coverage. We selected the dust producing
    S stars in the ISO/SWS archive with enough signal to perform a
    detailed dust analysis, and then compare the dust spectra from the
    9 sources with the O-rich AGB spectra and a subset of M
    super-giants. We constructed average dust emission spectra of the
    different categories.}
  {We report the discovery of several previously unreported dust
    emission features in the S star spectra. The long wavelength
    spectra of W~Aql and $\pi^1$~Gru exhibit the ``30''~$\mu$m feature
    attributed to MgS. Two sources exhibit a series of emission bands
    between 20 and 40~$\mu$m that we tentatively ascribe to Diopside.
    We show that the 10$-$20~$\mu$m spectra of the S stars are
    significantly different from the O-rich AGB stars. The O-rich
    stars exhibit a structured emission feature that is believed to
    arise from amorphous silicate and aluminium-oxide. The S stars
    lack the substructure found in the O-rich stars. Instead they show
    a smooth peak with a varying peak-position from source to source.
    We suggest that this feature is caused by a family of related
    material, whose exact composition determines the peak position.
    The observed trend mimics the laboratory trend of
    non-stoichiometric silicates. In this scenario the degree of
    non-stoichiometry is related to the Mg to SiO$_4$ ratio, in other
    words, to the amount of free O available during the dust grain
    growth.}
  {}

\maketitle

\section{Introduction}
\label{sec:intro}
\begin{table*}
  \caption{Details of the sources/SWS spectra used in this study.}
  {\small
  \begin{tabular}{l@\  l l l l@\  l l@\  l}
    \hline
    \hline
    Source& 
    IRAS name&
    AOT$^a$&
    TDT$^b$& 
    Class.$^c$&
    SP98$^d$&
    Spec.Type$^e$& 
    Remarks
    \\
    \hline
    \multicolumn{8}{c}{S stars from \citet{1994yCat.3168....0S} used in
    this study}\\
    \hline
    R And      & 00213+3817 & 01(2) & 40201723 & 2.SEc  & SE3 & S3,5-8,8e       & 18~$\mu$m feat. weak\\
    S Cas      & 01159+7220 & 01(2) & 41602133 & 3.SEp  & SE3 & S3,4-5,8e       & 15~$\mu$m feat., 18~$\mu$m feat. absent, C$_2$H$_2$+HCN abs.\\
    W Aql      & 19126-0708 & 01(2) & 16402335 & 3.SEp  & SE3 & S3,9-6,9e       & 15~$\mu$m feat., 18~$\mu$m feat. absent, C$_2$H$_2$+HCN abs.\\
    R Cyg      & 19354+5005 & 01(1) & 42201625 & 2.SEb  & SE3 & S2.5,9-6,9e     & 18~$\mu$m feat. absent, 10~$\mu$m feat. narrow\\
    $\chi$ Cyg & 19486+3247 & 01(2) & 15900437 & 2.SEb  & SE3 & S6,2-10,4e      & 18~$\mu$m feat. absent, substructure at 10.2~$\mu$m\\
    AA Cyg     & 20026+3640 & 01(2) & 36401817 & 2.M    & N   & S7,5-7.5,6      & :            \\              
    RZ Sgr     & 20120-4433 & 01(2) & 14100818 & 2.SEa  & SE2 & S4,4ep          & \\
    $\pi^1$~Gru& 22196-4612 & 01(2) & 34402039 & 2.SEa  & SE2 & S7,5e           & 18~$\mu$m feat. absent \\
    RX Lac     & 22476+4047 & 01(1) & 78200427 & 2.SEa  & SE1 & M7.5Se          & :\\
    \hline
    \multicolumn{8}{c}{Stars from \citet{1994yCat.3168....0S} not included in the S star sample}\\
    \hline
    W Cet      & 23595-1457 & 01(2) & 37802225 & 2.SEa: & -   & S7$^{e1}$       & Low SNR\\                    
    T Cet      & 00192-2020 & 01(2) & 55502308 & 2.SEa  & SE1t& M5-6S IIe       & M supergiant, M, MS or S\\   
    RW And     & 00445+3224 & 01(3) & 42301901 & 7      & SE3:& M5-10e(S6,2e)   & Low SNR\\                    
    WX Cam     & 03452+5301 & 01(2) & 81002721 & 1.NO   & -   & S5/5.5$^{e2}$   & Low SNR\\                    
    NO Aur     & 05374+3153 & 01(1) & 86603434 & 2.SEa  & SE1 & M2S Iab         & Supergiant\\                 
    LY Mus     & 13372-7136 & 01(2) & 13201304 & 1.NO   & -   & M4III$^{e1}$    & No dust features\\           
    II Lup     & 15954-5114 & 06    & 29700401 & 3.CE$^{c1}$&-& SC$^{e3}$       & Carbon-rich\\                
    ST Her     & 15492+4837 & 01(3) & 41901305 & 2.SEa  & SE1 & M6-7 IIIaS      & MS\\                         
    OP Her$^f$ & 17553+4521 & 01(1) & 77800625 & 1.NO   & N   & M5 IIb-IIIa(S)  & No dust features\\           
    HD 165774  & 18058-3658 & 01(2) & 14100603 & 1.NO:  & -   & M2II/III$^{e1}$ & No dust features, low SNR\\  
    S Lyr      & 19111+2555 & 01(1) & 52000546 & 2.CE:  & SE2 & SCe             & Low SNR\\                    
    HR Peg     & 22521+1640 & 01(2) & 37401910 & 1.NO   & -   & S5,1$^{e1}$     & No dust features\\           
    GZ Peg     & 23070+0824 & 01(3) & 37600306 & 1.NO   & N:  & M4S III         & No dust features\\           
    \hline
    \multicolumn{8}{c}{Supergiants with 10~$\mu$m features resembling the S stars}\\
    \hline
    KK Per     & 02068+5619 & 01(1) & 45701204 & 2.SEa  & -   & M2 Iab          & 18~$\mu$m feat. weak, UIR\\
    V605 Cas   & 02167+5926 & 01(2) & 61301202 & 2.SEa: & -   & M2 Iab$^{e4}$   & 18~$\mu$m feat. weak, UIR \\
    AD Per     & 02169+5645 & 01(2) & 78800921 & 2.SEap & -   & M2.5 Iab$^{e5}$ & 18~$\mu$m feat. weak, UIR\\
    NO Aur     & 05374+3153 & 01(1) & 86603434 & 2.SEa  & SE1 & M2S Iab         & 18~$\mu$m feat. weak, UIR \\
    V1749 Cyg  & 20193+3527 & 01(2) & 73000622 & 2.SEb  & SE3t& M3 Iab          & UIR\\
    IRC+40 427 & 20296+4028 & 01(3) & 53000406 & 2.SEap:& -   & M0-2 I$^{e6}$   & 18~$\mu$m feat. absent, UIR\\
    CIT 11     & 20377+3901 & 01(1) & 40503119 & 2.SEb  & -   & M3: Iab         & 18~$\mu$m feat. weak\\
    V354 Cep   & 22317+5838 & 01(2) & 41300101 & 2.SEc  & SE6 & M2.7 Iab        & 15~$\mu$m feat., 18~$\mu$m feat. absent\\
    V582 Cas   & 23278+6000 & 01(1) & 38501620 & 2.SEc  & SE5 & M4 I$^{e7}$     & 18~$\mu$m feat. weak, UIR:\\
    \hline
    \multicolumn{8}{c}{Other sources used in this study}\\
    \hline
    IRC+50 096 & 03229+4721 & 01(2) & 81002351 & 3.CE   & -   & C$^{e2}$        & Carbon-rich, ``30''~$\mu$m feature \\
    R Hya      & 13269-2301 & 01(1) & 08200502 & 2.SEa  & SE2t& M6-9eS(Tc)      & \\
    TY Dra     & 17361+5746 & 01(2) & 74102309 & 2.SEc  & SE8t& M5-8            & \\
    \hline
  \end{tabular}}\\
$^{a}$Observing mode used \citep[see][]{1996A&A...315L..49D,
  1996A&A...315L..38C}. Numbers in brackets correspond to the
scanning speed. 
    $^{b}$TDT number which uniquely identifies each ISO observation.
    $^c$Classification of the SWS spectrum from
    \citet{2002ApJS..140..389K}, except
    $^{c1}$\citet{2003ApJS..147..379S}. 
    $^d$IR classification in the scheme of
    \citet{1998ApJS..119..141S}. 
    $^e$Spectral types are from \citet{1998ApJS..119..141S}, except
    $^{e1}$\citet{2001KFNT...17..409K},
    $^{e2}$\citet{2001yCat.3222....0B},
    $^{e3}$\citet{1998yCat.3206....0B},
    $^{e4}$\citet{1970AJ.....75..602H},
    $^{e5}$\citet{1955ApJ...122..434B},
    $^{e6}$\citet{1978A&AS...34..409S} and
    $^{e7}$\citet{1994AAS...185.4515W}.
    $^f$This source is only listed in the first edition of general
    catalogue of S stars \citep{1995yCat.3060....0S} and is classed M
    star in the second edition.
  \label{tab:smooth_sources}
\end{table*}
\begin{figure*}
  \sidecaption
  \centering \includegraphics[clip,width=12cm]{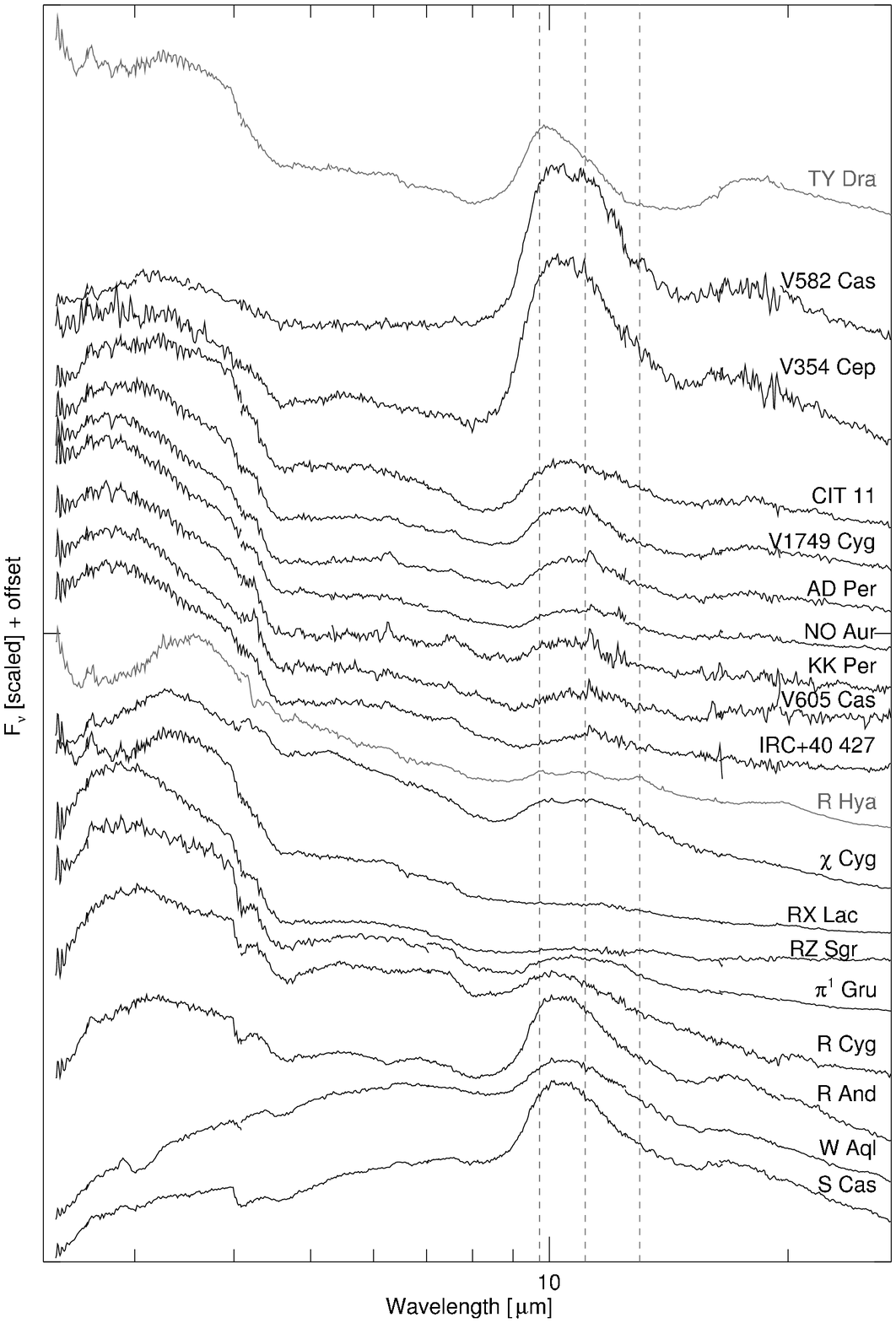}
  \caption{ISO/SWS spectra of the sources listed in
    Table~\ref{tab:smooth_sources}. From the bottom to the top we
    show. Eight spectra of S stars (S~Cas $-$ $\chi$~Cyg). The grey
    spectrum in the middle is R Hya, an O-rich AGB star that shows the
    typical structured 10~$\mu$m feature. Then nine M super-giants
    that exhibit a smooth and displaced 10~$\mu$m features (IRC+40~427
    $-$ V582~Cas). At the very top we show TY~Dra as an example of
    classical silicate dominated spectrum. The dashed lines represent
    the wavelengths of the substructures found in the O-rich AGB
    stars.}
  \label{fig:smoothies}
\end{figure*}
Asymptotic giant branch (AGB) stars are evolved stars of low to
intermediate mass in the ZAMS mass range of $\sim$1 to 8~M$_\odot$.
During the AGB phase, these stars are typified by high luminosity and
low surface temperature, which implies a very large radius and low
surface gravity. These stars often exhibit substantial mass loss
through a dust driven or pulsation driven wind. This mass loss is
important for several reasons. \emph{i)} These winds provide the means
through which these stars return nucleosynthesis products (i.e.
metals) from the interior of the star into the interstellar medium
(ISM). As the bulk of all stars fall in this mass range, the winds of
AGB stars are one of the dominant contributors to the enrichment of
the ISM. \emph{ii)} These dusty winds alter the appearance of these
stars because they cause the star to be surrounded by an envelope of
gas and dust. The dust will absorb the stellar radiation and reradiate
in the IR; therefore if one is to study the properties of these stars,
the dusty envelope needs to be taken into account. \emph{iii)} The
further evolution of the star is determined by the mass loss. Unlike
most types of stars, AGB star evolution is not determined by the
nuclear fusion processes in the interior but by the mass loss at the
surface and the AGB will end when the reservoir of envelope material
has been exhausted by the mass loss.

The type of molecules and solid-state particles that are present in
the winds are to the first order determined by the elemental
abundances at the surface of the star. Most important in this respect
is the number ratio of C-atoms to O-atoms (C/O ratio), because the
carbon-monoxide (CO) molecule is easily formed and very stable. This
causes most of the C- and O-atoms to be present in the form of CO.
These atoms are then effectively not taking part in the chemistry and
dust- condensation that occurs in these surroundings. Only the
fraction of either O-atoms if C/O is smaller than unity, or C-atoms if
C/O exceeds unity, is available. This dichotomy as a function of C/O
ratio is clearly found in the AGB stars. When C/O$<$1 , i.e. in O-rich
AGB stars, one finds large amounts of oxygen-bearing molecules, like
SiO, H$_2$O and CO$_2$, while for C/O$>$1, i.e. in C-rich AGB stars,
molecules like CH, C$_2$H$_2$ and HCN are present. The same holds for
the composition of the dust around these objects. The O-rich stars
exhibit silicates and oxides like amorphous aluminium-oxide
(Al$_2$O$_3$) or spinel, while their C-rich counterparts produce
amorphous carbon, silicon carbide and sulfides like MgS.

During the AGB phase the elemental abundances on the surface of the
star are being altered by a process called dredge-up. This causes
fusion products from the interior to be transported to the surface,
gradually increasing the C/O ratio. So, in the broadest possible terms
stars on the AGB gradually evolve from O-rich AGB stars to C-rich AGB
stars. In this scenario the S stars form an interesting intermediate
class of objects that have C/O$\simeq$1. Simplemindedly, one would
assume that in such surroundings the range of molecules and dust
components could be very wide perhaps overlapping with both the
typical O-rich and C-rich species or species that are not found in
either surroundings. Theoretically, iron-silicide and metallic iron
are the dust species predicted on the basis on chemical equilibrium
calculations \citep{2002A&A...382..256F}. Observationally, this class
of objects has not been exhaustively studied to determine the dust
composition although some studies have focused on the 10~$\mu$m
emission feature from S stars \citep{1988ApJ...333..305L,
  1993ApJ...416..769C, 1998ApJS..119..141S, 2000A&AS..146..437S}. In
particular the spectra obtained with the Short Wavelength Spectrometer
(SWS) \citep{1996A&A...315L..49D} on-board the Infrared Space
Observatory (ISO) \citep{1996A&A...315L..27K} which cover a much
broader wavelength range (2$-$45~$\mu$m) than available before with a
high sensitivity and spectral resolving power, allow us to get a more
complete picture of the composition around S stars.

The most diagnostic feature in the S star spectra is found in the
10~$\mu$m region. The dust emission features in the 10~$\mu$m region
are very extensively studied because \emph{i}) this window is
available for ground-based observations, \emph{ii}) it holds the
important diagnostic resonances of both O-rich (silicates, Al$_2$O$_3$
and spinel) and C-rich (SiC) dust and \emph{iii}) spectra for many
evolved stars in this region are available in the form of IRAS/LRS
spectra. Many studies have focused on classifying the spectral
appearance of the $\sim$10~$\mu$m emission band, using various
classification schemes, and relating the classification to the other
observable characteristic of the sources like the spectral type,
luminosity class, the C/O ratio, the variability type or the mass-loss
rate \citep{1988ApJ...333..305L,1990MNRAS.243...78S,
  1990AJ.....99.1173L,1993ApJ...416..769C,1995ApJ...451..758S,
  1998ApJS..119..141S,1999MNRAS.309..180S,2000A&AS..146..437S,
  2000ApJ...531..917M,2002ApJS..140..389K,2007A&A...463..663Y}. Here
we do not repeat the details of these studies but focus on the main
conclusions that are relevant for the current work and refer to
\citet{1998ApJS..119..141S} and \citet{2000A&AS..146..437S} for
extensive and thorough comparisons of the relevant literature and the
various interpretations of the dust emission characteristics.

The main points that have emerged are as follows:
\begin{enumerate}
\item O-rich AGB stars show a range of dust emission features with a
  range of peak positions.
  \begin{enumerate}
  \item A single narrow peak at 9.7~$\mu$m (on a F$_\nu$ scale), due to
    ``classical silicates''. The corresponding Si-O bending mode
    resonance is also observed near 18~$\mu$m.
  \item A broader structured feature, with local maxima at 9.7, 11 and
    13~$\mu$m, with the peak-position close to $\sim$10~$\mu$m.
  \item A broad and shifted feature peaking at longer wavelength (up
    to 13~$\mu$m). This broad feature exhibits the same local maxima
    as mentioned above. It is generally accepted that this broad
    feature is due to the dominant contribution of dust components
    other than the silicates, probably amorphous aluminium-oxide
    \citep{1989A&A...218..169O}.
  \end{enumerate}
  The origin of the structured features (b) in between the two
  extremes above is uncertain. While \citet{1989A&A...218..169O}and
  \citet{2000MNRAS.315..856L} model these spectra with a mixture of
  silicates and aluminium-oxide, \citet{2001ApJ...558..165E} find that
  these observations, with the additional constraints set by the IRAS
  broadband photometry, are best explained using pure silicates in a
  dust shell of larger optical depth. However, the fact that one finds
  the same substructures within a wide range of peak-positions argues
  against the opacity effect being dominant, and most likely these
  sources represent a mixture of silicate and aluminium-oxide grains.
\item M super-giants exhibit a similar range of 10~$\mu$m features,
  although they are more weighted to the classical silicate profiles
  (a) and the broad (c)-type features are rare
  \citep{1998ApJS..119..141S}. An interesting exception is a
  significant fraction of M super-giants of the h and $\chi$~Per
  association that exhibit a (c)-type feature peaking near 10.5~$\mu$m
  with the 11.3~$\mu$m UIR band perched on top
  \citep{1998MNRAS.301.1083S}. In Sect.~\ref{sec:discussion} we will
  compare these interesting sources with the S stars.
\item \citet{1988ApJ...333..305L} find that S-type stars
  preferentially exhibit a broad feature peaking in the
  10.5$-$10.8~$\mu$m range, which in fact makes them dub this feature
  the ``S'' feature. This agrees well with the findings of
  \citet{1998ApJS..119..141S}, who also find that the S-type stars in
  their sample predominantly exhibit such a broad feature. In contrast
  \citet{1993ApJ...416..769C} report that the S-type stars have
  emission features very similar to M stars, covering the complete
  range from pure silicates to the broad feature profile. This
  difference is easily understood from the fact that the latter
  authors include in their much larger sample also many MS and weak S
  stars which have a C/O ratio well below unity and should have dust
  similar to the M stars. The profile found in the strong S stars is
  broad and peaking in the 10.5$-$10.8~$\mu$m range. The extend and
  peak-position of this S star feature are very similar to the broad
  features found in the O-rich AGB stars, however the 13~$\mu$m
  feature is lacking in the S star spectra
  \citep[][Fig.9]{1998ApJS..119..141S}. These authors also find that
  there might be a slight enhancement between 10 to 11~$\mu$m in the
  S-type feature when compared to the M star sample, although they
  state that the difference might not be significant given the quality
  of the LRS spectra they used.
\end{enumerate}

In this paper we explore the dust composition around S stars as
derived from the available SWS spectra.

\section{Observations}
\label{sec:observations}
The ISO/SWS database contains 22 S star spectra that cover the full
wavelength range of 2.3-45~$\mu$m: 21 ISO/SWS sources listed in the
second edition of the general catalogue of S stars
\citep{1994yCat.3168....0S} with the addition of the S star OP~Her are
given in Table~\ref{tab:smooth_sources}. Of these 22 sources we
include nine in the S star sample that we explore here; NO~Aur is a
super-giant, II~Lup is carbon-rich, ST~Her is an MS star and T~Cet is
of uncertain nature. The other stars have either no significant dust
emission or too little signal-to-noise to investigate their dust
composition to the level of detail required here.

\subsection{Data reduction}
The data were processed using SWS interactive analysis, IA
\citep[see][]{1996A&A...315L..49D}, using calibration files and
procedures equivalent to pipeline version 10.1. Further data
processing consisted of extensive bad data removal and rebinning on a
fixed resolution ($\lambda$/$\Delta\lambda$=200) wavelength grid. In
order to combine the different sub-bands into one continuous spectrum
from 2 to 45 $\mu$m we applied scaling factors. In general the match
between the different sub-bands is good and the applied
scaling/offsets are small compared to the flux calibration
uncertainties with a few exceptions:

\subsubsection{$\chi$ Cyg }
The band 2A (4.1$-$5.3~$\mu$m) and 2C (7.0$-$12.5~$\mu$m) data are
affected by strong memory effects. The shapes of band 2A of the up and
down scans differ substantially. The slope of the band 2C data differs
while the details of the shape and substructure are present in both
scans. The data appear to be also slightly affected by miss-pointing
as we have to apply a scaling of $\sim$1.08 to the band 1
(2.4$-$4.1~$\mu$m) data and 1.2 to the band 3 (12.5$-$29~$\mu$m) data.

\subsubsection{$\pi^1$~Gru}
\begin{figure}
  \centering \includegraphics[clip,width=8.8cm]{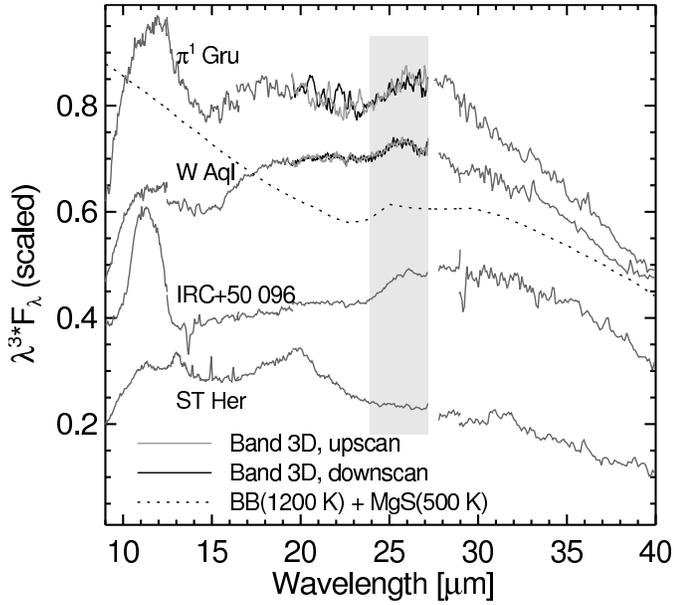}
  \caption{Comparison of the features found in the SWS spectra of
    $\pi^1$~Gru and W~Aql to the MgS feature in IRC+50~096. We show
    the signal of the two independent scans of sub-band 3D. The
    characteristic sharp rise from 24 to 26 (indicated in the light
    shaded area) with a gradual decline towards longer wavelength is
    found in all available scans. The scale along the ordinate is
    chosen to better bring out the structure in this wavelength domain
    without having to resort to removing an underlying baseline. The
    dotted line shows the spectral signature expected from a star
    surrounded by MgS, simulated with a Planck-function of 1200~K plus
    the emission of MgS grains at 500~K in a CDE-shape distribution.
    Near the bottom we show the spectrum of the MS star ST~Her as an
    example of a source that does not show this feature.}
  \label{fig:updown}
\end{figure}

The data for $\pi^1$~Gru are apparently affected by a slight
miss-pointing and the data for the sub-bands 2A to 3D (4.1 $-$ 27.5
$\mu$m) need to be multiplied by a relatively large factor
($\sim$1.25) to be consistent with the flux levels at shorter and
longer wavelengths. We note that the main signature of the MgS feature
is contained within sub-band 3D (19.5 $-$ 27~$\mu$m) and is not
affected by the scaling (see also Fig.~\ref{fig:updown}).

The final reduced spectra are presented in Fig.~\ref{fig:smoothies}.
As it turns out these stars are very rich in their dust emission
spectra exhibiting a wide range of dust features from different types
of materials. We will first discuss the long wavelength part of the
spectra of \object{$\pi^1$ Gru}, \object{W Aql} and \object{RX Lac} as
they show features that have not been detected before around S stars.

\section{Long wavelength spectra}
\begin{figure}
  \centering
  \includegraphics[width=8.8cm]{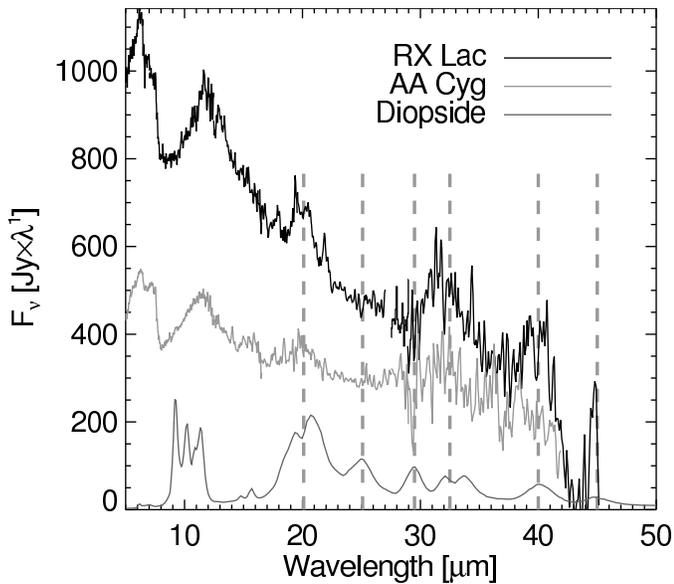}
  \caption{Long wavelength spectra of RX~Lac and AA~Cyg. The intensity
    units are chosen to better exhibit the emission bands on the
    steeply dropping stellar continuum. Below we show the absorption
    cross-sections of Diopside
    \citep[MgCaSi$_2$O$_6$,][]{2000A&A...363.1115K}}
  \label{fig:rx_lac}
\end{figure}

\subsection{The ``30 $\mu$m'' feature}
In Fig.~\ref{fig:spec_overview} we show the spectra of $\pi^1$~Gru and
W~Aql.  We also show for comparison the SWS spectrum of the C-rich red
giant \object{IRC+50~096}, which exhibits purely C-rich dust features
and the MS star \object{ST~Her}, which exhibits O-rich dust features.

The most remarkable feature in the spectra of these two S stars is the
emission feature starting at 23.5~$\mu$m and peaking at 26~$\mu$m,
indicated in Fig.~\ref{fig:spec_overview} with MgS. The feature is
weak and even in the closeup of the region in the right panel of
Fig.~\ref{fig:spec_details} not very prominent. This is probably also
the reason why it has eluded detection until now.

In Fig.~\ref{fig:updown} we display the spectra in intensity units
that better reveal the structure in this region. The shape of the
feature found in the S star spectra closely corresponds to the shape
of the much stronger MgS resonance found in the C-rich star
IRC+50~096. In particular the change of slope from 22 to 24 $\mu$m and
the second change beyond 26 $\mu$m is detected in the three sources at
the top. We also show the expected spectrum from MgS, which exhibits
the same shape. The two independent scans in the SWS data yield the
same shape, attesting to the reality of the MgS detection.

The presence of MgS around these stars is unexpected and prompts
several questions on dust condensation conditions around such stars.
It is important to also take into account other aspects of these
systems. First we will discuss the other dust features in the spectra
of these stars and subsequently the molecular composition.

\subsection{RX Lac and AA Cyg}
In Fig.~\ref{fig:rx_lac} we show the spectra of RX~Lac and AA~Cyg.
Both sources exhibit structure at 20, 32.5 and 40~$\mu$m. The
structure around 20~$\mu$m seems to be quite secure in both spectra.
The long wavelength part of the AA~Cyg spectrum is very noisy.
Comparing the detected features with available laboratory spectra they
seem to correspond best to Diopside. There might be a connection
between the appearance near 10~$\mu$m as both sources exhibit a weak,
very red emission feature and strong SiO absorption. This absorption,
in combination with a low Diopside temperature, might explain why we
do not detect the corresponding 10~$\mu$m bands of Diopside. In should
be borne in mind that the long wavelength spectra of these stars are
quite noisy and that the correspondence with the laboratory spectrum
is not perfect. In particular the 25~$\mu$m feature which is present
in the laboratory spectrum is markedly absent in the stellar
emissions. Therefore this should be considered a tentative
identification.

\section{The broad feature in the SWS spectra of S stars}
\begin{figure}
  \centering \includegraphics[clip,width=8.8cm]{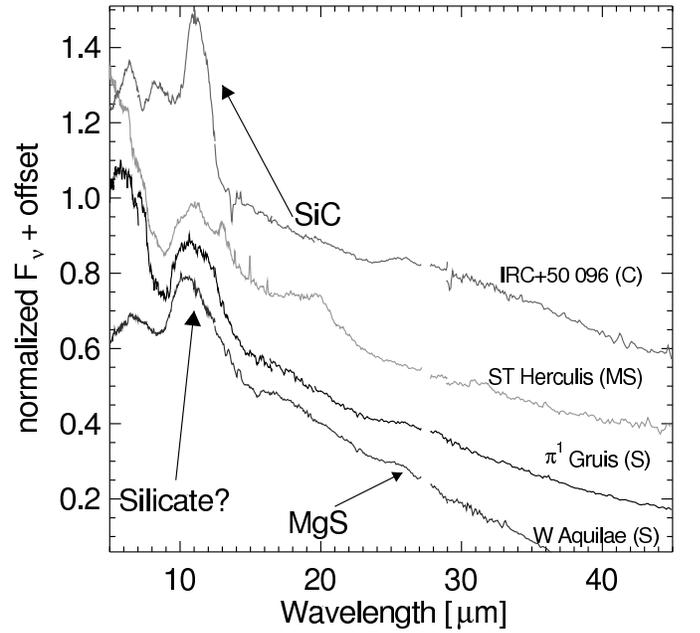}
  \caption{Overview of the SWS spectra of four red giant stars. We
    show from top to bottom the carbon-rich AGB star IRC+50~096, the
    M6.5S star ST~Her and the S stars $\pi^1$~Gru and W~Aql. The
    carbon-rich star exhibits the typical SiC feature at 11.3 $\mu$m
    and the MgS feature near 26 $\mu$m. The ST~Her spectrum is
    dominated by the silicate dust features at 10 and 20 $\mu$m which
    are typical for oxygen-rich environments. $\pi^1$~Gru and W~Aql
    exhibit the MgS feature (see also Fig.~\ref{fig:spec_details} \&
    \ref{fig:updown}) and a broad feature around 10 $\mu$m which
    resembles the silicate emission but is lacking clear evidence of
    the corresponding 20~$\mu$m silicate band. The latter is
    especially clear in the spectrum of $\pi^1$~Gru, while there is a
    weak feature near $\sim$17~$\mu$m in the spectrum of W~Aql.}
  \label{fig:spec_overview}
\end{figure}
\begin{figure*}
  \centering \includegraphics[clip,height=18cm,angle=90]{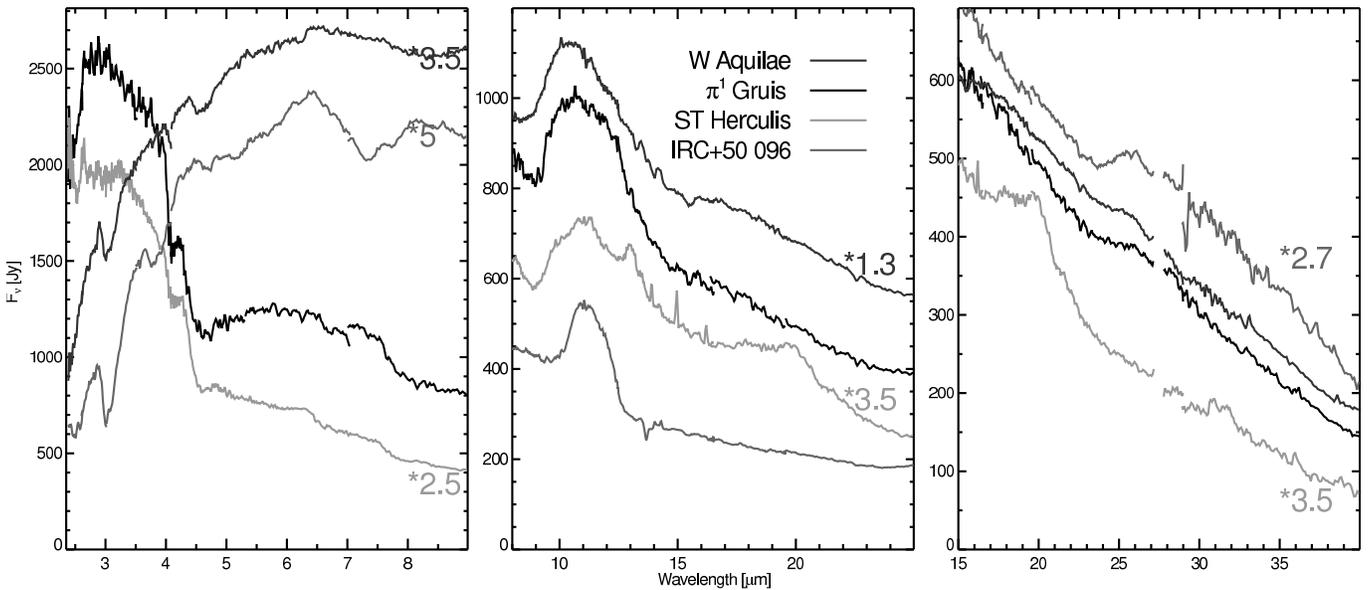}
  \caption{Detailed view of the IR spectra of the stars presented in
    Fig.~\ref{fig:spec_overview}. The left panel shows the shortest
    wavelength region in which the spectral structure is dominated by
    molecular absorption bands. Clearly the molecular composition of
    $\pi^1$~Gru resembles that of ST~Her. The typical C-rich molecular
    bands due to C$_2$H$_2$, HCN and C$_3$ are absent in $\pi^1$~Gru
    but are seen in the IRC+50~096 and W~Aql. The middle panel
    compares the 10 to 20 $\mu$m range of these sources. The
    wavelength range of the broad $\sim$10~$\mu$m emission band in the
    S star spectra is identical to the silicate emission of ST~Her,
    however the feature appears more smooth and the sharp
    substructures at 9.5, 10.8 and 13 $\mu$m are absent. Likewise, the
    20~$\mu$m silicate band is not observed. The silicon-carbide feature
    seen in IRC+50~096 corresponds in peak position but is much
    narrower than the emission feature in ST~Her. Finally, the panel
    on the right shows the MgS emission feature which is detected in
    the IR spectra of $\pi^1$~Gru, W~Aql and IRC+50~096 and not in
    ST~Her.}
  \label{fig:spec_details}
\end{figure*}
\begin{figure}
  \includegraphics[width=8.8cm]{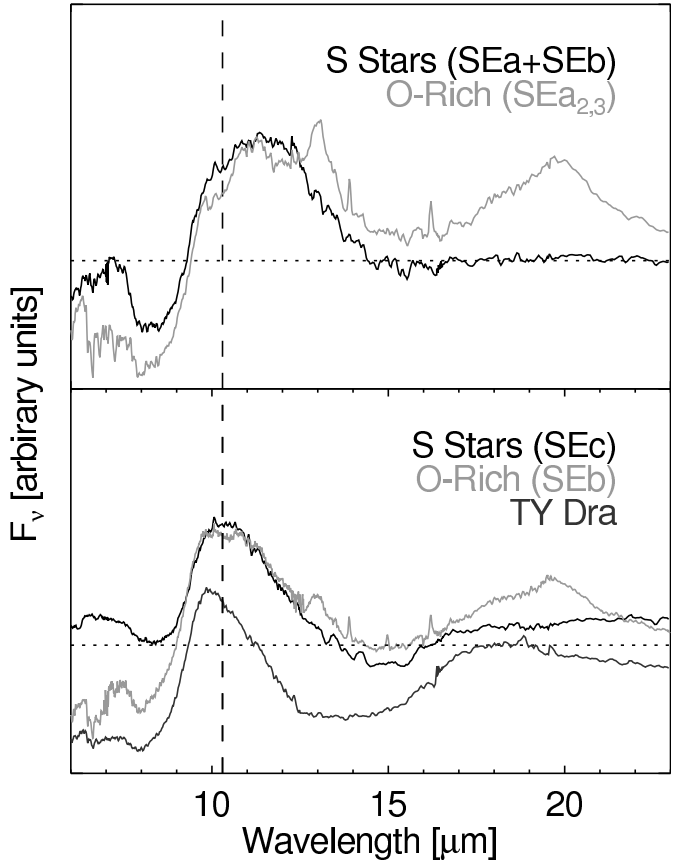}
  \caption{Comparison of the mean profiles of the S stars and the
    O-rich stars. The samples have been split in two parts. Those with
    profiles peaking beyond 10.5~$\mu$m (top panel) and those that
    peak at shorter wavelengths (bottom panel). The dashed line
    indicates the position (10.3~$\mu$m) where the mean profiles
    differ significantly. For reference we also show the feature
    spectrum of TY~Dra, which shows a classical silicate emission
    feature. The spectrum has been offset for clarity.}
  \label{fig:mean_profiles}
\end{figure}
\begin{figure}
  \includegraphics[width=8.8cm]{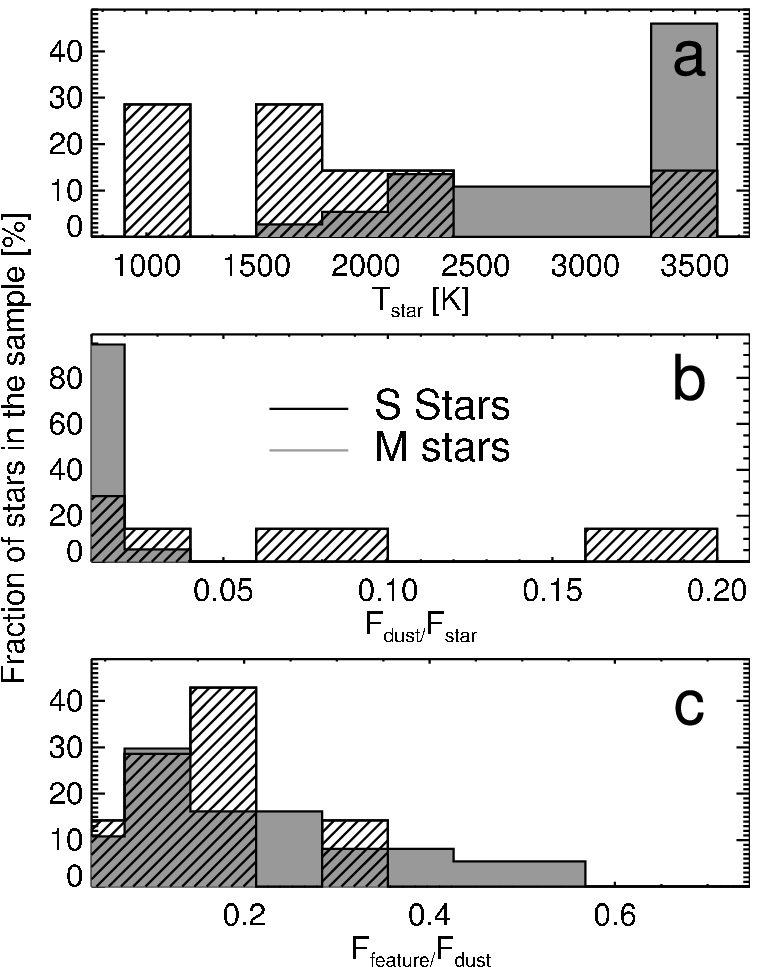}
  \caption{Properties of the M and S stars. From top to bottom: the
    fitted temperature of the (reddened) stars; the ratio of the
    excess to the stellar flux and the ratio of the flux in the
    feature to the flux in the dust excess. The S stars in the sample
    are clearly cooler (i.e. more extincted) and show more excess
    than the M stars.}
  \label{fig:measurements}
\end{figure}
The most prominent dust feature in the spectra of these S stars is the
broad and smooth emission feature centred around $\sim$11~$\mu$m
(Fig.~\ref{fig:spec_overview} and the middle panel of
Fig.~\ref{fig:spec_details}). A similar feature (covering the same
wavelength range) is found in the IR spectra of many O-rich evolved
stars
\citep[e.g.][]{1968ApJ...154..677G,1990AJ.....99.1173L,2002ApJS..140..389K},
which it is usually attributed to a combination of silicates and
aluminium-oxide with the possible addition of spinel at 13~$\mu$m
\citep[e.g.][]{1969ApJ...155L.181W,1972A&A....21..239H,1999A&A...352..609P,Cami_PhD}.
However, the corresponding silicate feature due to the Si-O
stretch/bend near 18~$\mu$m is remarkably weak.

The similarities between the S star 10~$\mu$m features and the broad
feature found in the O-rich spectra have until now been interpreted as
the S stars showing silicate plus aluminium-oxide emission like the
O-rich stars. In the following we will argue against this
interpretation. Instead, we will show that there is a significant
difference between the M star broad features due to silicates plus
aluminium-oxide and the S star broad features, pointing to a different
dust composition for these S stars. In the following we concentrate on
comparing the S star spectra with O-rich stars that exhibit low dust
columns, since these O-rich stars show the structured 10~$\mu$m
emission complex. The stars that have higher mass-loss rates show
profiles close to the classical silicate emission
\citep[e.g.][]{2005A&A...439..171H}. See TY~Dra in
Fig.~\ref{fig:smoothies} for an example of such a classical silicate
feature.

We readdress the issue of the spectral appearance of the 8$-$22~$\mu$m
dust spectra of the S stars using the available SWS spectra. We choose
to restrict ourselves to the relatively small sample of SWS spectra
for several reasons. The first reason is to have a consistent data
set. More importantly, the sensitivity and spectral resolving power
allow us to discuss details not available in the LRS spectra or in
many of the ground-based spectra. As an example we mention \object{R
  Aql} for which \citet{2000A&AS..146..437S} report a broad feature
without detectable substructure based on their CGS3 spectra, while the
SWS spectra clearly resolve the substructures at 9.7, 11 and
13~$\mu$m. Finally, the much wider wavelength coverage allows us to
better separate the different molecular and dust contributions. For
example, the 10~$\mu$m region feature of the Mira variable \object{RR
  Per} resembles the S star spectra. However, the complete SWS
spectrum reveals very strong molecular absorption and emission bands.
This spectrum could have been misinterpreted, in case only the 8 $-$
22~$\mu$m range would had been available.

\citet{2002ApJS..140..389K} have classified the ISO/SWS spectra
according to the shape of the continuum and dust features. Within
their classification $\pi^1$~Gru, W~Aql and many O-rich stars belong
to the same class exhibiting the broad feature due to silicate and
aluminium-oxide. Although the peak position and wavelength extend of
the emission bump are indeed very similar, the O-rich sources show
more substructure with often a prominent sharp emission maximum at
13~$\mu$m and always substructure at 9.7 and $\sim$11~$\mu$m
\citep[e.g.][ see also Fig.~\ref{fig:spec_details}]{Cami_PhD}. This
substructure is not found in the spectra of $\pi^1$~Gru and W~Aql.

A careful survey of all evolved star spectra with sufficient signal to
noise in the SWS database yields 9 more sources with a broadened and
smooth $\sim$10~$\mu$m emission feature. These sources are listed in
Table~\ref{tab:smooth_sources} and their spectra are shown in
Fig.~\ref{fig:smoothies}. The main conclusion we draw is that such a
broad and smooth emission feature is only found in the M super-giant
and the S star spectra, \emph{there are no O-rich AGB stars that
  exhibit the same emission feature}. This agrees very well with the
findings of \citet{2000A&AS..146..437S} that the broad feature in the
AGB stars differs from the broad features found in the super-giants.
The ISO spectra show that in fact the broad O-rich AGB star features
are always due to a mixture of the 9.7, 11 and 13~$\mu$m bands in
which the 11~$\mu$m band dominates. Admittedly the substructures are
sometimes more pronounced than in other cases which leaves open the
possibility that part of the emission in this region, even in the
O-rich AGB stars spectra, is due to the same smooth feature present in
the S stars. $\chi$~Cyg is the only S star that has exhibits these
substructures although the predominant contribution might still be a
smooth underlying feature.

Note that the sample of M super-giants displayed in
Fig.~\ref{fig:smoothies} largely overlaps with the broad featured M
super-giants presented by \citet{2000A&AS..146..437S} and consists
mostly of super-giants located in the h and $\chi$~Per association.
These sources are by no means typical for all galactic M super-giants.
As we already pointed out, the large majority of studied M
super-giants exhibits a classical silicate emission feature
\citep{1998ApJS..119..141S,2000A&AS..146..437S}. Moreover, seven out
of the nine broad featured M super-giants in our sample reveal UIR
emission bands, which are not commonly found in the M super-giant
spectra \citep{1994MNRAS.266..640S,1998MNRAS.301.1083S}.
\citet{1998MNRAS.301.1083S} suggest that this could be related to
peculiar elemental abundances at the stellar surfaces.

Another striking aspect of these spectra as a group is the relative
strength of the 10~$\mu$m feature compared to the 18~$\mu$m feature.
In the spectra of the O-rich stars, the 10~$\mu$m emission is
accompanied by an emission feature near 18$-$20~$\mu$m, where one
would expect the Si-O bending modes of the silicates to be present
(see Fig.~\ref{fig:spec_details}). This emission band is
systematically weak in the M super-giant and S star spectra in
Fig.~\ref{fig:smoothies}. In some of the sources we find no evidence
of the 18~$\mu$m band at all and three sources exhibit a weak emission
band near 15~$\mu$m instead.

\subsection{Feature extraction}
The differences in the observed features as discussed above are
relatively subtle. In order to enhance the contrast we have removed
the ``continuum'' from the spectra. We stress that the continuum as
such does not have any clear physical meaning in this context as the
same material(s) that give rise to the features contribute to this
continuum. It represents only a means to remove those contributions
that do not give rise to spectral signatures. It further allows us to
compare the relative strength of the remaining bands to other
components. Because of the wide range of properties of the stars in
the sample -- in terms of optical depth, strength of the molecular
bands and the excess -- we have opted to use a crude method for
continuum estimation. The continuum is represented by a sum of a warm
black-body-function (F$_\nu$ = B$_\nu$(T); the reddened stellar
photosphere) and a cooler modified black-body-function (F$_\nu$ =
$\nu^{p}~\times~$B$_\nu$(T); the dust excess). These two functions are
fitted simultaneously to selected ``continuum-points''. The latter are
difficult to define unambiguously and we choose to use the following
ranges: 2.9$-$3.35~$\mu$m, when the molecular absorption is not too
prominent; 3.35$-$3.8~$\mu$m; 7$-$7.5$^\dagger$~$\mu$m;
8.5$-$9.5$^\dagger$~$\mu$m; 14$-$15~$\mu$m; 22$-$25~$\mu$m and
36$-$44~$\mu$m. The ranges marked with a dagger often do not exhibit
clear continuum character. These regions have still been included,
albeit with a reduced weight, to account for the fact that the
continuum should run close to them.

The resulting residual spectra have been co-added to reduce the noise.
The co-addition has been applied in two bins, depending on the
peak-position of the ``10''~$\mu$m emission feature. The results are
shown in Fig.~\ref{fig:mean_profiles}, where we compare the S star
profiles to the features found in O-rich stars. Note that the
comparison is done on the basis of the extracted feature and does not
follow strictly the silicate-index classification. We find that both
S-star SEa and SEb classes compare best with the O-rich SEa class and
the S-star SEc class with the O-rich SEb class. This mismatch reflects
the fact that the underlying continua differs systematically -- the
O-rich stars being bluer -- between both groups of sources (see also
below).

Fig.~\ref{fig:mean_profiles} demonstrates that indeed the O-rich star
exhibit a ``10''~$\mu$m feature which is composed of three distinct
components while the S star spectra exhibit only a single broad
emission band -- indicative of a different dust composition between
the two groups. There are several other ``properties'' of these dust
spectra that change in league with the differences in dust
composition. The lack of the 19.5~$\mu$m band in the S star spectra is
pronounced. In Fig.~\ref{fig:measurements} we summarise the derived
properties of the S stars compared to those of the M stars. The main
conclusion is that the studied S stars as a sample are redder and
exhibit a stronger excess than the O-rich stars with the broad
10~$\mu$m feature. This indicates the presence of more dust along the
line of sight towards the S stars. There is no significant difference
in the strength of the 10~$\mu$m feature relative to the dust
continuum (Fig.~\ref{fig:measurements}c).

\section{Discussion}
\label{sec:discussion}
As discussed above, the spectral appearance of S stars is
significantly different from the O-rich AGB stars. Here we explore the
possible explanations for these differences, that is, the relative
strength of the emission bands and the displacement of the 10~$\mu$m
feature. The amount of mid-IR excess is probably not directly related
to the dust composition. Perhaps the observed difference in dust
composition and the differences in the quantities of dust share a
common origin in the particular condition that prevail during the S
star evolutionary state.

There are several effects that, if at work, will influence the
spectral appearance of dust features even arising from grains with the
same chemical composition. The most notable are dust temperature and
grain-size. Using the amorphous olivine
(Mg$_{0.8}$Fe$_{1.2}$SiO$_{4}$) that explains (part of) the emission
of the O-rich stars well \citep{2005A&A...439..171H}), we find that
the temperature required to explain the observed S star
10-to-18~$\mu$m ratio (ignoring the mismatch in peak-position) needs
to be higher than 2000~K. And even at such high temperatures (well
above the evaporation temperature of silicates) the feature at
18~$\mu$m would still be too prominent for the most extreme cases,
e.g. \object{$\pi^1$ Gru} or \object{W Aql}. Thus, dust temperature is
excluded as the cause of the weak 18-20~$\mu$m features.

Grain-size: We have simulated the effects of grain-size on the
spectral features of silicates based on simulated optical properties
using a Mie calculation \citep{BohrenHuffman}. The effect of
increasing the grain-size from 0.01 to 1 $\mu$m is indeed a
displacement of the 10~$\mu$m emission feature, but the effect is
subtle and it moves the peak by less than 0.3~$\mu$m. This is not
enough to explain the S star spectra. Moreover the ratio of the 10 to
18~$\mu$m band is only slightly affected. The 18~$\mu$m band does not
shift significantly but becomes somewhat stronger. The S star
spectra exhibit an 18~$\mu$m feature which is little pronounced and
peaks at shorter wavelength. We conclude that the grain-size is not
the dominant factor for explaining the S star silicate feature.

The paper by \citet{2003A&A...408..193J} is very interesting in terms
of the compositional influence on the silicate emission. These authors
have studied the IR transmission spectra of non-stoichiometric
magnesium-rich silicates. They find that there is a significant shift
of the 10~$\mu$m Si-O stretching resonance as a function of the Mg to
SiO$_4$ ratio. The trend is such that as the Mg to SiO$_4$ ratio
increases, the 10~$\mu$m peak shifts to longer wavelength, the
18~$\mu$m band broadens and weakens compared to the continuum and
shifts to shorter wavelength. This mimics to a large extent the
behaviour observed in the S star spectra. It should be noted that
although the 10~$\mu$m feature in the laboratory spectra shifts by a
large amount ($>$1~$\mu$m), it is not completely sufficient to explain
the full range observed in the stellar spectra. If the observed shift
is indeed linked to the presence of non-stoichiometric silicates, than
there could be an obvious connection to the photospheric abundances
since the Mg to SiO$_4$ ratio in the silicate might measure the Mg to
free O or SiO in the gas phase. In case of C/O $\sim$ 1 there is more
Mg in relative terms and those non-stoichiometric silicates may form.

Moreover, the M-super-giants that we present are special not just in
their 10~$\mu$m spectrum but also because they exhibit a dual
chemistry with both silicates and PAHs. How this dual chemistry comes
about is at present unclear. This may be evidence of the presence of a
disk plus wind geometry, shocked regions in the outflow or an
abundance pattern in the outflow, which permit this dual chemistry to
exist. In any case, it is clear that in some region of the
circumstellar environment the chemistry should be close to a C/O
around unity.

It should be stressed that we cannot draw general conclusions about
the dust formation around S stars from such a small number of sources.
We have successfully proposed to observe 90 S stars with IRS
\cite{2004AAS...204.3304H} on Spitzer (GO-30737, PI. Hony). The main
findings of the current paper, i.e., the presence of the MgS feature
and the shifted 10~$\mu$m feature, are borne out by the spectra
obtained in the Spitzer sample (Smolders et al, in prep). Because of
the larger sample the range of spectral features, in particular in the
10-20~$\mu$m region is larger than what is presented here.

\section{Conclusions}
We have presented a detailed study of dust spectra of the S stars
observed by ISO/SWS. The spectra exhibit several unique dust
characteristics. In particular, we find two S stars ($\pi^1$~Gru and
W~Aql) that exhibit a weak ``30''~$\mu$m feature due to MgS grains. RX
Lac and \object{AA Cyg} exhibit features between 20$-$40~$\mu$m which
may be related to Diopside, although this is at most a tentative
identification. The 10~$\mu$m region of the dust-producing S stars
stands out due to its smooth and broad emission feature. The feature
is located clearly at longer wavelengths than the classical silicate
feature, without showing the characteristic substructure found in the
spectra of the O-rich AGB stars with a broad 10~$\mu$m feature. The
peculiar 10~$\mu$m features of the S stars are accompanied by very
weak or absent features near 18~$\mu$m. We conclude that S stars make
different types of dust than the O-rich AGB stars, including those
that exhibit shifted 10~$\mu$m features. As a group the S stars and
their dust shells do not compare well to the O-rich AGB stars with
structured features, the S stars are redder and produce more dust. The
common explanation for the shifted 10~$\mu$m band, i.e. a mixture of
classical silicate with aluminium-oxide, does not appear to apply to
the S stars. We have explored possible origins of the peculiar
spectra. We find that non-stoichiometric silicates with an increased
Mg to SiO$_4$ ratio might be at the origin of the displaced emission.
Interestingly, the properties of the peculiar 10 and 20~$\mu$m
features are shared with a small subgroup of M super-giants. These
super-giants, which are preferentially found in the h and $\chi$ Per
associations, also exhibit the UIR emission features, again
strengthening the possible link with the abundances in the dust
forming regions.

\begin{acknowledgements}
  IA$^3$ is a joint development of the SWS consortium. Contributing
  institutes are SRON, MPE, KUL and the ESA Astrophysics Division.
  This work was supported by the Dutch ISO Data Analysis
  Center(DIDAC). The DIDAC is sponsored by SRON, ECAB, ASTRON and the
  universities of Amsterdam, Groningen, Leiden and Leuven.
\end{acknowledgements}

\bibliographystyle{aa}
\bibliography{articles}
\end{document}